\documentstyle[preprint,aps]{revtex}

\begin{document}
\title{Scattering of Hot Excitons in Quantum Wells}
\author{Meng Lu$^{(a)}$, Joseph L. Birman$^{(a)}$, Carlos Trallero-Giner$^{(b)}$, and
Fernando de Le\'{o}n-P\'{e}rez$^{(b,c)}$}
\address{(a) Department of Physics, The City College of CUNY, New York, NY 10031\\
(b) Department of Theoretical Physics, Havana University, 10400 Vedado,
Havana, Cuba\\
(c) Physics Department, Central University ``Marta Abreu'' of Las Villas,
Santa Clara, Cuba}
\date{\today}
\maketitle

\begin{abstract}
The scattering probabilities of hot excitons in narrow quantum wells (QWs)
are obtained. The exciton-phonon matrix element is considered by using an
envelope function Hamiltonian approach in the strong quantization limit
where the QW width is smaller than the exciton bulk Bohr radius. The Fr\"{o}%
hlich-like interaction is taken into account and the contribution of the
confined and interface modes to the scattering probability are calculated as
a function of quantum well width, electron and hole effective masses, and
in-plane center-of-mass kinetic energy. Inter- and intra-subband excitonic
transitions are discussed in term of the phonon scalar potential selection
rules. It is shown that even parity electrostatic potential states for
confined and interface modes give the main contribution to the excitonic
scattering rate. The consequences of exciton relaxation and
scattering probability present for the multiphonon resonance Raman
scattering in narrow QWs are briefly discussed.
\end{abstract}

\pacs{PACS: ?}

\section{ Introduction}

In this work we investigate the hot exciton scattering rates in
semiconductor narrow quantum wells (QWs). An exciton with kinetic energy
higher than the thermal temperature of the medium is called a hot exciton.
This concept has been introduced in the past when multiphonon LO-lines have
been observed in resonant Raman scattering in bulk semiconductor materials
(see \cite{Gross} and references therein). For polar semiconductors when the
exciton energy is greater than a longitudinal optical phonon, the scattering
rate is governed by the exciton-LO phonon interaction and the Fr\"{o}hlich
like mechanism is the stronger one playing the main role in the exciton
relaxation time.

In the past and for the bulk semiconductor case, transition probabilities of
hot excitons interacting with longitudinal and acoustic phonons have been
carried out in Refs. [2-4]. The dependence on exciton kinetic energy,
effective electron and hole masses, and on the excitonic quantum numbers were
obtained, showing the conditions that favored the hot excitons contribution
to the multiphonon Raman process in semiconductors.

The importance of hot exciton relaxation via optical phonon emission in QWs
has been stressed by photoluminescence experiments \cite{Sapega} and by
time-resolved Raman scattering measurements \cite{Thatan}. Also, the
relative role of confined phonons and interface modes in time resolved
resonant Raman spectroscopy experiments is pointed out in Ref. \cite{Tsen}.
Today, there are several theoretical studies of carrier relaxation times in
QWs considering the main mechanism due to electron-LO-phonon interaction
(e.g. Ref. \cite{Lee} and references therein). Nevertheless, a direct
calculation of the hot exciton scattering rate due to the confined optical
phonons in a QW is lacking in the literature.

In this paper we will focus on the excitonic scattering rate calculation and
on its contribution to the multiphonon resonant Raman process (MPR) in
narrow QWs. Mowbray et al \cite{Mowbray} reported a
multiphonon process in GaAs/AlAs superlattices. Raman spectra for the second
and third order processes from GaAs phonons show structures involving the
combination of confined and interface LO phonons. The relative
intensities of the second and third order processes due to the combination
of $L_{2}$ confined phonon are weaker than that obtained with $L_{2}$ and 
interface phonons for samples of 6-GaAs monolayer. The enumeration of phonon symmetries in
the QW follows Ref. \cite{Trallero}. For the case of 10-GaAs monolayer
samples, the combined intensities of $L_{2}$ and $L_{4}$ \ phonons are
comparable to the intensities due to $2L_{2}$ or $3L_{2}$ ones. These
results show that the relative intensities of a MPR scattering process are
very sensitive to the quantum well width. The connection of the MPR process
with excitonic transitions followed when the excitonic cascade model is
invoked to explain the main features of the scattered light \cite{Gross}.
According to this model, the scattering cross section with an emission of
one emitted phonon is proportional to the exciton scattering rate $W(E)$
with kinetic energy $E$. Thus, an explicit calculation of $W(E)$ as a
function of the QW width and phonon mode contribution will bring an
further understanding of MPR process in QWs as we will discuss 
bellow.

\section{Scattering probability}

The scattering probability per unit time of an exciton in a QW is given by
the Fermi ``golden rule''

\begin{equation}
W=\frac{2\pi }{\hbar }\sum_{f}|M|^{2}\delta (E_{f}-E_{i}),  \label{probab}
\end{equation}
where $E_{i}$ and $E_{f}$ are the energies at the initial and final states,
$M$ is the matrix element describing the scattering of the hot exciton. For
the case we are dealing with, the transitions are due to the interaction
between excitons and LO-phonons. In a QW the interaction Hamiltonian is
given by

\begin{equation}
H_{int}=\sum_{n_{p},{\bf q}}\bar{C}_{F}\left[ \phi _{n_{p},{\bf q}}(z_{e})%
\mathop{\rm \mbox{{\Large e}}}\nolimits^{i{\bf q}.{\bf \rho _{e}}}-\phi
_{n_{p},{\bf q}}(z_{h})\mathop{\rm \mbox{{\Large e}}}\nolimits^{i{\bf q}.%
{\bf \rho _{h}}}\right] \hat{b}_{n_{p},{\bf q}}+H.c.\;\;,
\label{hamiltoniano}
\end{equation}
where $\hat{b}_{n_{p},{\bf q}}$ is the LO-phonon creation operator with
in-plane wavevector ${\bf q}$ and frequency $\omega _{n_{p},{\bf q}}$, ${\bf %
\rho _{e}}({\bf \rho _{h}})$ and $z_{e}(z_{h})$ are the electron (hole)
coordinates in the X-Y plane and along the growth direction, respectively,
and $\bar{C}_{F}$ is the Fr\"{o}hlich constant. To describe the confined
phonons and interface modes we follow the phenomenological treatment given
in Refs. \cite{Trallero,Comas}. The function $\phi _{n_{p},{\bf q}}$ shows a
definite parity \cite{Note} and the corresponding analytical expression for
phonon dispersion relation for odd and even potential states have been
considered in Ref. \cite{Comas}. The phonon dispersion for a structure of
GaAs/AlAs with 2.0 nm and 1.7 nm well width are shown in Fig. 1. The
branches are labeled as $L_{N}$ and $T_{N}$ ($N$=1,2,...) according to their
longitudinal and transverse character at ${\bf q}=0$. The modes which show a
strong dispersion as a function of the in-plane wavevector ${\bf q}$ are
those with the strongest electric field component and also the strongest
interface character. For the 2.0 nm QW the two interface branches are $L_{1}$
and $L_{4}$ modes, while in the case of Fig. 1(b) these corresponds to $L_{1}
$ and $T_{1}$ modes. The upper mode is purely antisymmetric ($L_{1}$) while
the lower interface mode ($L_{4}$ and $T_{1}$ in Figs. 1(a) and (b),
respectively) is symmetric. The antisymmetric part of the potential ($L_{1}$%
, $L_{3}$ modes) couples to the exciton via inter-subband transitions with
exciton wavefunctions having different symmetry with respect to the center
of the well. The symmetric part of the function $\phi _{n_{p},{\bf q}}$ ($%
L_{2}$, $L_{4}$ modes) couples to excitons via intra-subband transitions
since the exciton wave function have the same symmetry respect to Z-axis.
The exciton inter-subband transition is also possible through the symmetric
phonon potential part, if the initial and final excitonic states have the
same parity with respect to the center of the QW.

For the excitonic states we assume a Wannier-Mott model in the quantum well
in the strong quantization limit (very narrow QWs). In this case, the QW
width is smaller than the exciton Bohr radius $a_{B}$ (d 
\mbox{$<$}%
\mbox{$<$}%
$a_{B}$) and the three-dimensional Wannier exciton motion can be reduced to
two-dimensional Wannier exciton and one-dimensional electron and hole motion
along the growth direction \cite{MengLu}. Hence, the wave function $\Psi
_{exc}({\bf r}_{e},{\bf r}_{h})$ can be cast in the form

\begin{equation}
\Psi _{exc}({\bf r}_{e},{\bf r}_{h})=\Psi _{N,m}({\bf R},{\bf \rho })\Phi
_{n_{e}}(z_{e})\Phi _{n_{h}}(z_{h}),  \label{fonda}
\end{equation}
where ${\bf R}$ is the electron-hole pair center-of-mass coordinates on the
X-Y plane, ${\bf \rho }$ is the in-plane relative coordinates. The electron
and hole motion along the Z-axis is described by $\Phi _{n_{e}}(z_{e})$ and $%
\Phi _{n_{h}}(z_{h})$ wavefunctions $(n_{e},n_{h}=1,2,...)$. The function $%
\Psi _{N,m}({\bf R},{\bf \rho })$ corresponds to the two-dimensional exciton
wavefunction with radial quantum number $N$ and angular quantum number $%
m=0,\pm 1,\pm 2,...$ with the bound state energies equal to

\begin{equation}
\epsilon _{N}=-\frac{R_{y}}{(N+1/2)^{2}}\;\;\;\;\;\;N=0,1,...
\end{equation}
where $R_{y}$ is the effective Rydberg constant. The total exciton energy is

\begin{equation}
E=E_{g}+E_{n_{e}}+E_{n_{h}}+\epsilon _{N}+\frac{\hbar ^{2}K^{2}}{2M}.
\label{Etotal}
\end{equation}
$E_{g}$ being the bulk gap energy, $E_{n_{e}}(E_{n_{h}})$ is the electron
(hole) energy in the QW along the Z-direction, and $\frac{\hbar ^{2}K^{2}}{2M%
}$ is the exciton kinetic energy of the center-of-mass in the X-Y plane. In
our case where $d<<a_{B}$ the exciton motion can be decoupled from the
Z-component and the three-dimensional Wannier exciton is taken in separable
form according to Eq. (\ref{fonda}) by the product of the wavefunction in
the plane $\Psi ({\bf R},{\bf \rho })$ and the electron and hole subbands $%
\Phi _{n_{i}}(z_{i})\;\;(i=e,h)$ wavefunctions. The 3D Coulomb interaction
effect on the total exciton energy (\ref{Etotal}) could be treated by
perturbation theory with the Hamiltonian $H_{p}=\frac{e^{2}}{\epsilon }\left[
1/|{\bf \rho }_{e}-{\bf \rho }_{h}|-1/|{\bf r}_{e}-{\bf r}_{h}|\right] $
(see Ref. \cite{Eduardo}).

Substituting Eqs. (\ref{fonda}) and (\ref{hamiltoniano}) in Eq. (\ref{probab}%
), one can find that the transition probability due to the interaction with
the phonon branch $n_{p}$ as a function of the exciton wave number $K$ is
given by

\begin{eqnarray}
&&W_{N,m;n_{e},n_{h}\rightarrow N^{\prime },m^{\prime };n_{e}^{\prime
},n_{h}^{\prime }}^{(n_{p})}=\alpha _{0}\int_{0}^{\infty
}dq\;q\int_{0}^{2\pi }d\theta _{1}\times   \nonumber \\
&&  \nonumber \\
&&\left| I_{N,m}^{N^{\prime },m^{\prime }}(Q_{h})<\Phi _{n_{e}}|\phi _{n_{p},%
{\bf q}}|\Phi _{n_{e^{\prime }}}>\delta _{n_{h},n_{h}^{\prime
}}-I_{N,m}^{N^{\prime },m^{\prime }}(Q_{e})<\Phi _{n_{h}}|\phi _{n_{p},{\bf q%
}}|\Phi _{n_{h^{\prime }}}>\delta _{n_{e},n_{e}^{\prime }}\right| ^{2}\times 
\nonumber \\
&&  \nonumber \\
&&\delta \left( \frac{\hbar ^{2}q^{2}}{2M}-\frac{\hbar ^{2}Kq}{M}cos\theta
_{1}+\hbar \omega _{n_{p},{\bf q}}+E_{n_{e^{\prime
}}}-E_{n_{e}}+E_{n_{h^{\prime }}}-E_{n_{h}}+\epsilon _{N^{\prime }}-\epsilon
_{N}\right) ,  \nonumber \\
&&  \label{W}
\end{eqnarray}
where $\theta _{1}$ is the angle between ${\bf K}$ and ${\bf q}$, $%
Q_{h(e)}=m_{h}(m_{e})a_{B}q/M$ with $m_{h}(m_{e})$ being the hole (electron)
mass, $\alpha _{0}=e^{2}\omega _{L}(\epsilon _{\infty }^{-1}-\epsilon
_{0}^{-1})/d$, $\omega _{L}$ is the LO bulk phonon frequency, and $\epsilon
_{\infty }(\epsilon _{0})$ is the high (static) frequency dielectric
constant. In the above expression only the phonon emission Stokes process
has been considered $(T\rightarrow 0K)$. It is clear that obtaining
scattering probability considering the phonon absorption process is
straightforward if the phonon occupation number $N({\bf q})=\left[ 
\mathop{\rm \mbox{{\Large
e}}}\nolimits^{\hbar \omega _{n_{p},{\bf q}}/k_{B}T}-1\right] ^{-1}$ is
introduced in the r.h.s. of Eq. (\ref{W}) and $\hbar \omega _{n_{p},{\bf q}}$
is substituted by $-\hbar \omega _{n_{p},{\bf q}}$. The first term in the
r.h.s of Eq. (\ref{W}) is due to the phonon emission by the electron with
matrix elements $I_{N,m}^{N^{\prime },m^{\prime }}(Q_{h})$ in the X-Y plane
and $<\Phi _{n_{e}}|\phi _{n_{p},{\bf q}}|\Phi _{n_{e}^{\prime }}>$ along
the Z-axis, while the second term is the hole contribution to the scattering
probability. The transition matrix elements $I_{N,m}^{N^{\prime },m^{\prime
}}$ and $<\Phi _{n}|\phi _{n_{p},{\bf q}}|\Phi _{n^{\prime }}>$ are given
elsewhere \cite{MengLu}. The energy and quasi-momentum conservation laws
impose restrictions on the wave number $q$ within the range $q_{min}\leq
q\leq q_{max}$, where $q_{min}$ and $q_{max}$ are solutions of the
equation

\begin{eqnarray}
f_{\pm }(E_{K}) &=&q^{2}\pm \frac{2}{a_{B}}\sqrt{\frac{M}{\mu }}\sqrt{\frac{%
E_{\kappa }}{R_{y}}}q  \nonumber \\
&+&\frac{M}{a_{B}^{2}\mu R_{y}}\left( \hbar \omega _{n_{p},{\bf q}%
}+E_{n_{e}^{\prime }}-E_{n_{e}}+E_{n_{h}^{\prime }}-E_{n_{h}}+\epsilon
_{N^{\prime }}-\epsilon _{N}\right) =0.  \label{fs}
\end{eqnarray}
Here, $\mu $ is the reduced mass, $E_{K}=\hbar ^{2}K^{2}/2M$ is the in-plane
exciton kinetic energy and the condition $q_{min}\geq 0$ must be fulfilled.

Integrating with respect to $\theta _{1}$ and summing over $m$ and $m^{\prime}$%
, we obtain from Eq. (\ref{W}) the transition probability

\begin{eqnarray}
W_{N;n_{e},n_{h}\rightarrow N^{\prime };n_{e}^{\prime },n_{h}^{\prime
}}^{(n_{p})}(E_{K}) &=&\alpha _{0}\frac{2M}{\hbar ^{2}}%
\int_{q_{min}}^{q_{max}}\frac{dq\;q}{\sqrt{f_{+}(E_{K})f_{-}(E_{K})}}\times 
\nonumber \\
&&  \label{Wf} \\
\sum_{m,m^{\prime }}|\mathop{\rm \mbox{{\Large e}}}\nolimits^{i|m-m^{\prime
}|\theta _{0}}I_{N,m}^{N^{\prime },m^{\prime }}(Q_{h}) &<&\Phi _{n_{e}}|\phi
_{n_{p},{\bf q}}|\Phi _{n_{e^{\prime }}}>\delta _{n_{h}^{\prime
},n_{h}}-I_{N,m}^{N^{\prime },m^{\prime }}(Q_{e})<\Phi _{n_{h}}|\phi _{n_{p},%
{\bf q}}|\Phi _{n_{h}^{\prime }}>\delta _{n_{e}^{\prime },n_{e}}|^{2} 
\nonumber
\end{eqnarray}
where

\begin{equation}
\tan (\theta _{0})=2\sqrt{f_{+}(E_{K})f_{-}(E_{K})}%
/(f_{+}(E_{K})+f_{-}(E_{K})).
\end{equation}
From the energy and quasi-momentum conservation laws it follows that the
scattering probability is different from zero if the exciton kinetic energy $%
E_{K}>\hbar \omega _{n_{p},{\bf q}}+E_{n_{e}^{\prime
}}-E_{n_{e}}+E_{n_{h}^{\prime }}-E_{n_{h}}+\epsilon _{N^{\prime }}-\epsilon
_{N}$.

In Fig. 2, we illustrate the content of Eq. (\ref{Wf}), showing the types
of one phonon Stokes scattering processes included in the scattering rate.
Each exciton band is characterized by single electron and hole quantum
numbers $(n_{e},n_{h})$ and the total quantum number $N$: two excitonic
bands are shown including dispersion due to center-of-mass motion as in 
Eq. (\ref{Etotal}). For an exciton excited to the initial state labeled A in 
Fig. 2 three processes are allowed in principle:

(a) intra-subband scattering, with phonon potential state of even parity $%
(n_{p}=2,4)$ and $n_{e}=n_{e}^{\prime },n_{h}=n_{h}^{\prime }$,

(b) inter-subband transition where $n_{e}\neq n_{e}^{\prime }$ and $%
n_{h}\neq n_{h}^{\prime }$ with even parity phonon potential if $%
|n_{e}-n_{e}^{\prime }|=even$ or/and $|n_{h}-n_{h}^{\prime }|=even$, and for
the odd parity potential case $(n_{p}=1,3)$ if $|n_{e}-n_{h}^{\prime }|=odd$
or/and $|n_{e}-n_{h}^{\prime }|=odd$, finally

(c) inter- intra-subband scattering with $n_{e}=n_{e}^{\prime }$ but $%
n_{h}\neq n_{h}^{\prime }$ and viceversa. Here, if $|n_{h}-n_{h}^{\prime }|$
is even the phonon potential is an even function, while for $%
|n_{h}-n_{h}^{\prime }|$ equal to an odd number, the phonon potential states
need to be odd giving a zero electron contribution to the scattering
probability.

However, in the approximation used in this paper, only two processes will
occur: (a) and (c), due to the separation of motion in the X-Y plane and the
orthogonal Z direction. This is illustrated in Fig. 2, where single phonon
inter- intra-subband scattering are shown by arrows A$\longrightarrow
B^{\prime },A\longrightarrow C^{\prime }$ defining the corresponding $q_{min}
$ and $q_{max}$ for theses processes.

The exciton intra-subband transitions are coupled to the symmetric part of
the phonon potential, while the exciton inter-subband scattering (with mixed
character or not) couples to the symmetric or antisymmetric part
of the Fr\"{o}hlich Hamiltonian interaction.

Hence, the total inverse exciton lifetime characterized by the quantum
numbers $N;n_{e},n_{h};E_{K}$ and due to the interaction with an optical
phonon $n_{p}$ can be written as

\begin{eqnarray}
W_{N;n_{e},n_{h}}^{(n_{p})}(E_{K}) &=& W_{N;n_{e},n_{h}\rightarrow 
N;n_{e},n_{h}}^{(n_{p})}(E_{K})+\sum_{N^{\prime }\neq N}
W_{N;n_{e},n_{h}\rightarrow N^{\prime };n_{e},n_{h}}^{(n_{p})}(E_{K})
\nonumber \\
&+&\sum_{n_{e}\neq n_{e}^{\prime},\text{or},n_{h}\neq n_{h}^{\prime},
N^{\prime}}W_{N;n_{e},n_{h}\rightarrow N^{\prime};n_{e}^{\prime},n_{h}^{\prime }}^{(n_{p})}(E_{K})  \label{Wq}
\end{eqnarray}

The first term in the r.h.s of Eq. (\ref{Wq}) represents the intra-band
scattering probability without changes in the internal state of the exciton
branches. The second term is the probability of the intra-band scattering
accompanied by transitions from $N$ to other internal quantum states $%
N^{\prime }$, while the last term gives the inter-subband contribution to
the exciton lifetime in the branch $n_{e},n_{h}$ with internal quantum
number $N$ and in-plane kinetic energy $E_{K}$.

In the next section we study the dependence of the scattering rate on the
in-plane exciton kinetic energy, on the phonon state $n_{p}$, and on the
electron and hole effective masses.

\section{Discussion and conclusions}

For the numerical calculation of the scattering rate given by Eq. (\ref{Wf})
we selected the GaAs/AlAs parameters given in \cite{Comas}. We used for the
masses the in-plane mass $m_{h\perp }=m_{0}/(\gamma _{1}\pm \gamma _{2})$
and along Z-direction $m_{hz}=m_{0}/(\gamma _{1}\mp 2\gamma _{2})$, where $%
\gamma _{1}$ and $\gamma _{2}$ are the Luttinger parameters \cite{Landolt}.
The sign (+) or (-) corresponds to the light or heavy character of the mass,
respectively. In Fig. 3 we show the dependence of the scattering rate on $%
E_{K}$ in the $n_{e}=1$, $n_{h}=1$ excitonic branch and for the excitonic
ground state $N=0\rightarrow N=0$, $W_{0\rightarrow 0}^{(n)}$. Fig. 3(a)
corresponds to a QW with a well width d=2 nm while Fig. 3(b) is for d=1.7
nm. The calculation represents the heavy hole contribution along the quantum
well growth direction (light hole character in the X-Y plane) due to the
GaAs-like phonon modes $L_{2}$ and $L_{4}$ ($n_{p}=2$ and $n_{p}=4$). Since
we are dealing with intra-band transitions ($n_{e}=n_{h}=1$ for the final
and initial states) the exciton-phonon interaction with odd modes is
forbidden. It can be seen in Fig. 3(a) (d=2 nm) that the main contribution
corresponds to the $L_{4}$ mode and it is almost 3 times stronger than the $%
L_{2}$ mode. Figure 3(b) shows opposite effect where the $L_{4}$ phonon mode
is one order of magnitude weaker than the $L_{2}$ one. To explain the above
result we must to go back to the phonon dispersion relation of Fig. 1. For
the d=2 nm QW the $L_{2}$ confined mode is almost flat while the $L_{4}$
phonon has a stronger dispersion as a function of the phonon wavenumber $q$
and it is in addition the symmetric QW phonon which has a predominantly
interface character. The electrostatic potential associated with the $L_{4}$
mode in Fig. 1(a) gives strong coupling with excitons by the Fr\"{o}hlich
interaction in intra-subband transitions. The modes with largest $\phi
_{n_{p},{\bf q}}$ component in Eq. (\ref{Wf}) are those which have the
largest interface contribution, explaining why the scattering rate assisted
by the $L_{4}$ phonon in Fig. 3(a) is stronger than that due to the $L_{2}$
confined mode. In the case of d=1.7 nm the $L_{4}$ GaAs-type phonon is more
confined and it moves to lower frequency with an almost flat dispersion
relation (see Fig. 1(b)). For the range of $q$ values where the phonons are
flat the Fr\"{o}hlich interaction is proportional to $1/\sqrt{q^{2}+q_{z}^{2}%
}$ with $q_{z}=\frac{n_{p}\pi }{d}$ ($n_{p}=1,2,...$) \cite{Trallero1}.
Hence, the contribution to the intra-band scattering rate of the $n_{p}=2$
mode is four times stronger than for the $n_{p}=4$.

Let us comment on the effect of the carrier effective masses on the
scattering rate. First, the confinement is a function of the hole masses
along the Z-axis; Second, the hole-phonon matrix element depends on the hole
masses through the $Q_{h}=m_{h}/Mqa_{B}$ factor and the range of values $%
q_{min}\leq q\leq q_{max}$ is a function of the electron-hole mass (in plane
along the Z-axis); Third, the scattering probability value strongly depends
on the in-plane density of states. Fig. 3(c) shows the intraband
exciton scattering rate for the case of the light-hole mass along the
Z-direction (heavy-hole character in X-Y plane) and d=2 nm. As in Fig. 3(a)
the $L_{4}$ mode contribution is stronger than the $L_{2}$, showing that the
relative intensity between several phonon branches is only a function of the
phonon dispersion which itself depends on the QW width. The scattering rate
values, in units of the constant $W_{0}$, for the light hole exciton
contribution is smaller than the heavy-hole exciton one (see Figs. 3(a) and
3(c)). It is important to note that for the case we consider of isotropic
effective masses and $m_{e}=m_{h}$ the intra-band transition $%
W_{0\rightarrow 0}^{(n_{p})}$ is identically zero \cite{Aristora}.

In conclusion we have performed an analysis of the exciton scattering rate
in narrow QWs. The relaxation of hot excitons accounts for the emission of
confined and interface optical longitudinal phonons due to the Fr\"{o}%
hlich-like exciton-phonon interaction. The results obtained above show a
different behavior of the scattering probability as a function of the well
width and in-plane exciton kinetic energy. The value of $W_{0\rightarrow
0}^{(n_{p})}$ is not equal to zero for $E_{K}>\hbar \omega _{n_{p},0}$ (for
the case of intra-subband transition), increases with increasing $E_{K}$
reaching a maximum and beginning to fall according to the law $E_{K}^{-1/2}$ 
\cite{Note1}. The relative strength of the different phonon modes is
proportional to the ratio of the square of their overlap integral of the
function $\phi _{n_{p},{\bf q}}$ with initial and final electron and hole
subband states in the GaAs/AlAs QW. For a given phonon state, inter-subband
transitions are less efficient processes on the exciton relaxation time in
comparison with the intraband probability.

The intra-subband scattering rate is accompanied only by symmetric phonon
states ($n_{p}=2,4...$) in correspondence with the observed multiphonon
Raman spectra in a series of short-period GaAs/AlAs superlattices \cite
{Mowbray}. The reported multiphonon peaks involving pure overtones of GaAs
phonons are combinations of the even confined modes (see Fig. 6 in Ref. \cite
{Mowbray}). For samples with 10 GaAs monolayer width (d $\approx $ 2 nm)
combination of $L_{2}$ and $L_{4}$ (such as $2L_{2}$, $2L_{4}$, $3L_{2}$, $%
L_{2}+2L_{4}$, etc) are observed in the GaAs phonon spectra. In the case of
samples with the 6 GaAs monolayer well width (d $\approx $ 1.7 nm) only the
combination of $L_{2}$ and interface GaAs modes ($2L_{2}$, $3L_{2}$, $L_{2}+I
$) are reported. Following the idea of the cascade model, where the iterated
exciton-one-phonon interaction is the dominant mechanism, the MPR
cross-section will be proportional to the $W_{0\rightarrow 0}^{(n_{p})}(E_{K})$
factor \cite{Trallero2} (the inter-band exciton transitions give very small
contribution and have been neglected). The obtained results shown on Fig.
3(b) predict that in a 6 monolayers sample, the MPR spectra needs to be
composed mostly by combinations of $L_{2}$ confined modes while the
combinations with $L_{4}$ must be almost forbidden. The opposite conclusion
is obtained from Fig. 3(a) (d=2 nm) when combinations of $L_{2}$ and $L_{4}$
GaAs phonon modes in the multiphonon Raman spectra should present similar
relative intensities.

Thus, the experimental observation in GaAs/AlAs superlattices on the
relative intensities of multiphonon scattering can be explained in the
framework of a MPR cascade model taking into account the relative role of
the confined and interface phonons on the hot exciton scattering rate.
Future work on the present issue is in progress.

\acknowledgments
This work was supported in part by a grant from the World Laboratory,
Lausanne. Support is also acknowledged from the Science Division of The City
College of CUNY and from the CUNY-Caribbean Exchange Program.

\newpage

{\bf Figure Caption}

Figure 1. Dispersion of the GaAs optical phonons of the 2.0 nm (a) and 1.7
nm (b) GaAs/AlAs quantum well. The phonons are labeled as longitudinal (L)
and transverse (T) according to its character at ${\bf q}=0$. Symmetric
phonon potentials correspond to $L_{2}$ and $L_{4}$, and antisymmetric
phonon potential to $L_{1}$ and $L_{3}$. The interface modes are those
showing strong dispersion as a function of $q$. The anticrossing of $L_{4}$
and $T_{1}$ modes near $q=0.5\times 10^{6}$ 1/cm is clearly seen in
figure (a). $L_{4}$ and $T_{1}$ modes present the same parity respect to
center of the QW.

Figure 2. Schematic representation of allowed scattering processes. See
text, section II, for discussion.

Figure 3. Exciton intra-subband scattering rate probability in units of $%
W_{0}=2\omega _{L}(\epsilon _{0}/\epsilon _{\infty }-1)dM/(a_{B}\mu )$ in a
narrow QW as a function of the in-plane center-of-mass exciton kinetic
energy. (a) and (b) correspond to the heavy-hole mass along the QW growth
direction with d=2 nm ($W_{0}=1.415\times 10^{12}$ 1/sec) and 1.7 nm ($W_{0}=1.203\times 10^{12}$ 1/sec), respectively. (c) represents the
light-hole mass along Z-direction for d=2 nm ($W_{0}=2.199\times 10^{12}$ 1/sec). The $L_{2}$ and $L_{4}$ phonon
modes according to the notation followed in Fig. 1 have been considered for
the one-phonon assisted exciton scattering rate.


\begin{references}
\bibitem{Note}  The electrostatic potential described in \cite{Comas}
present a misprint, a detailed study for the present problem is given in
Ref. \cite{MengLu}.
\end{references}
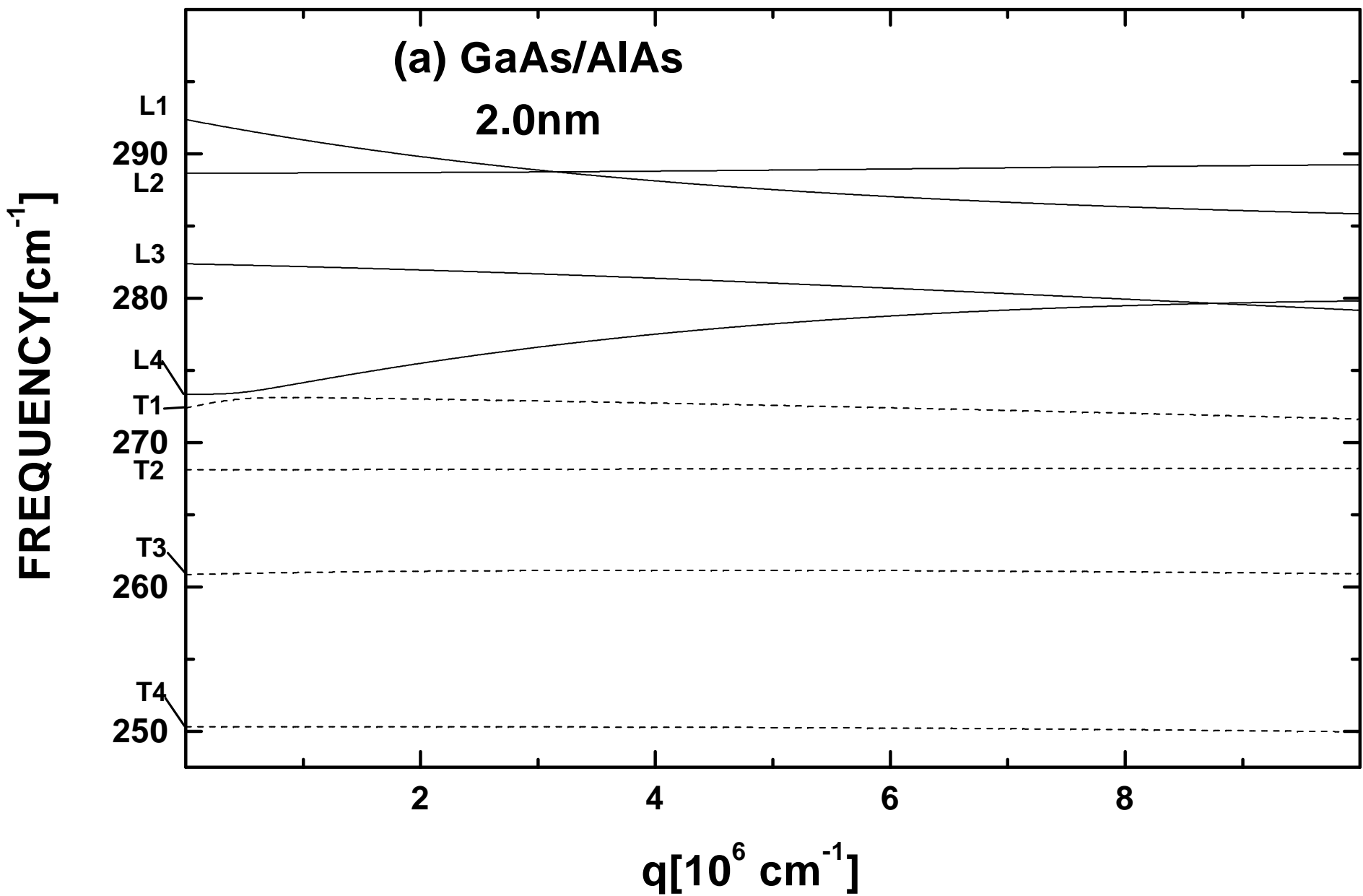
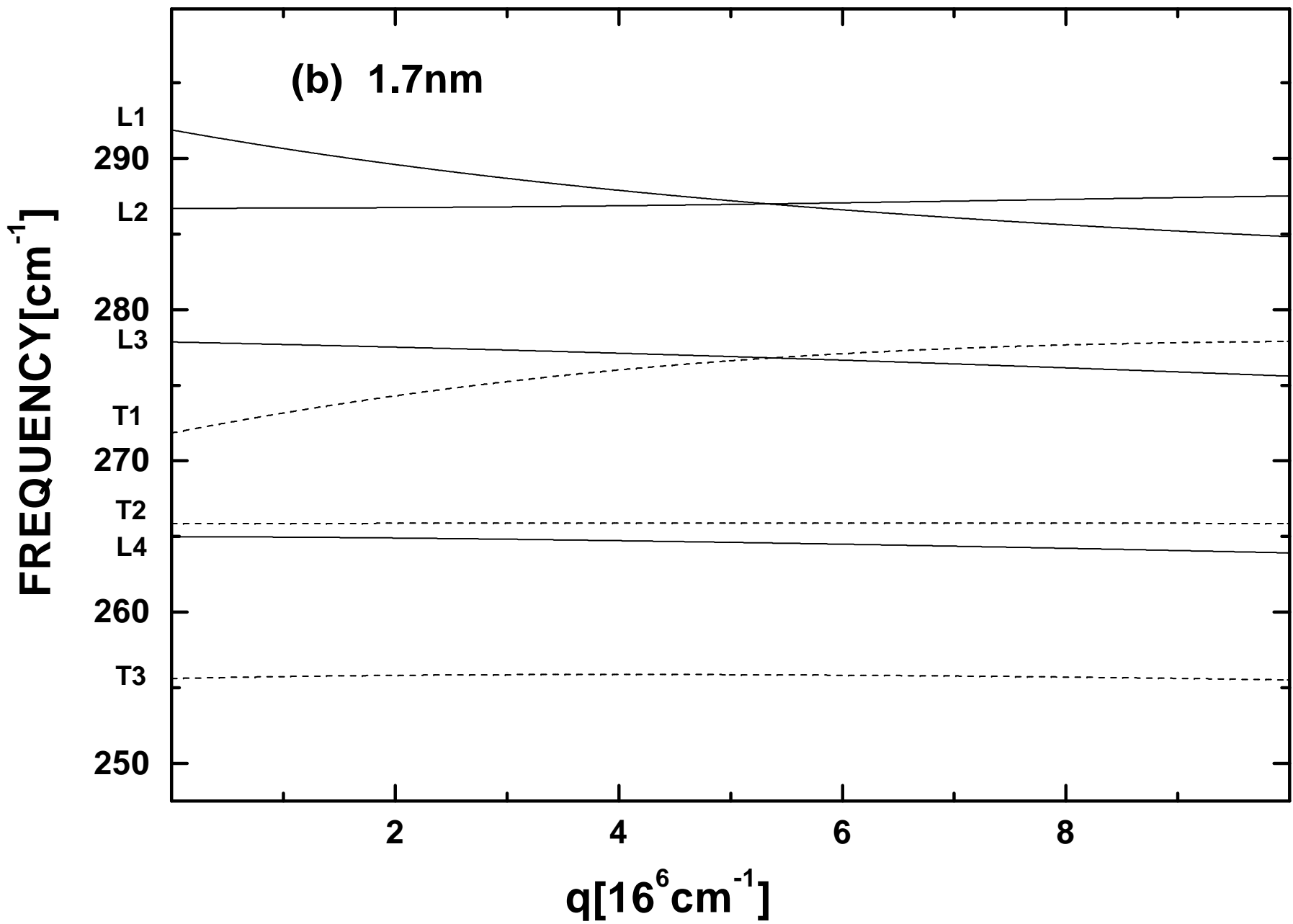
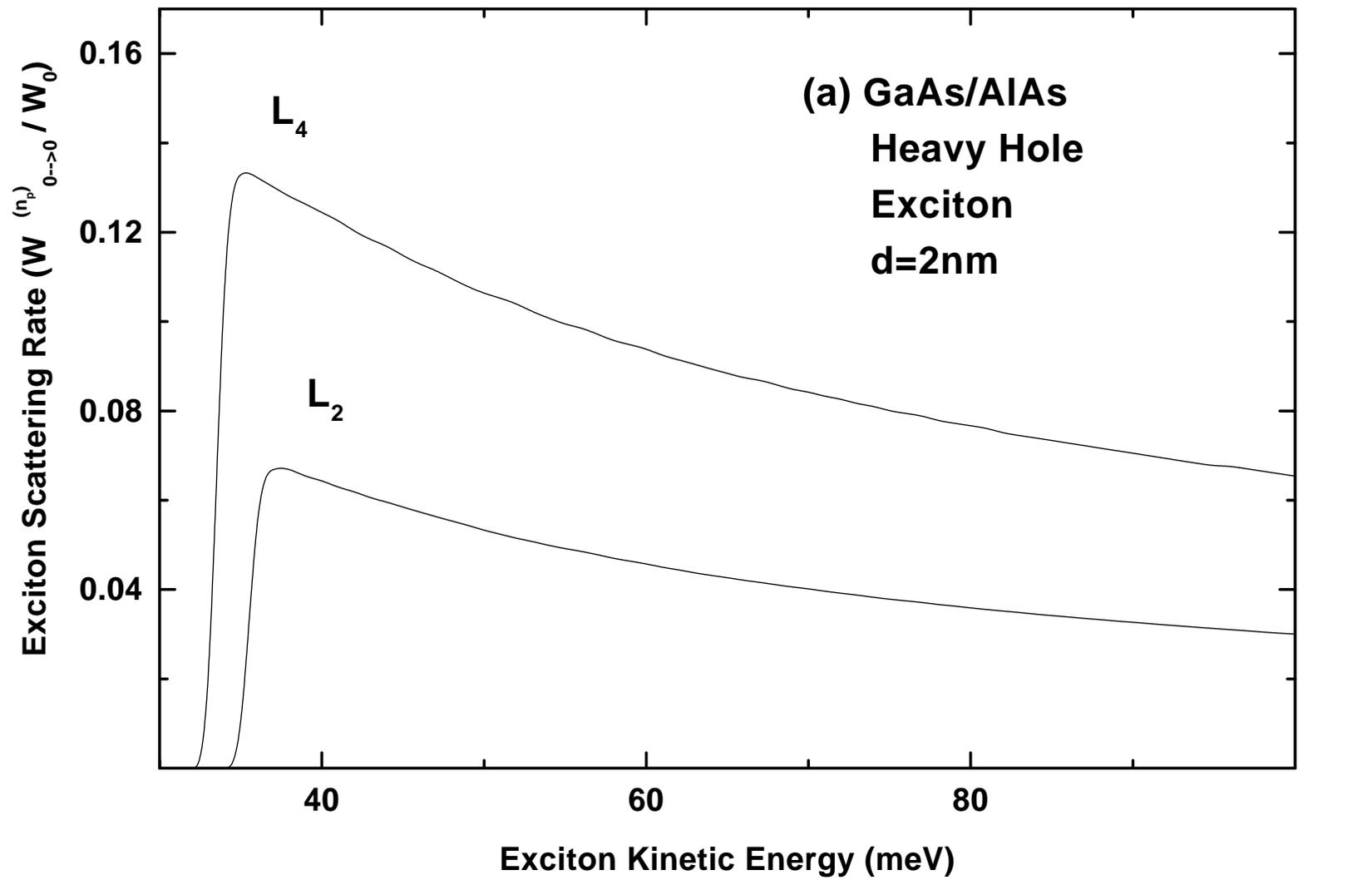
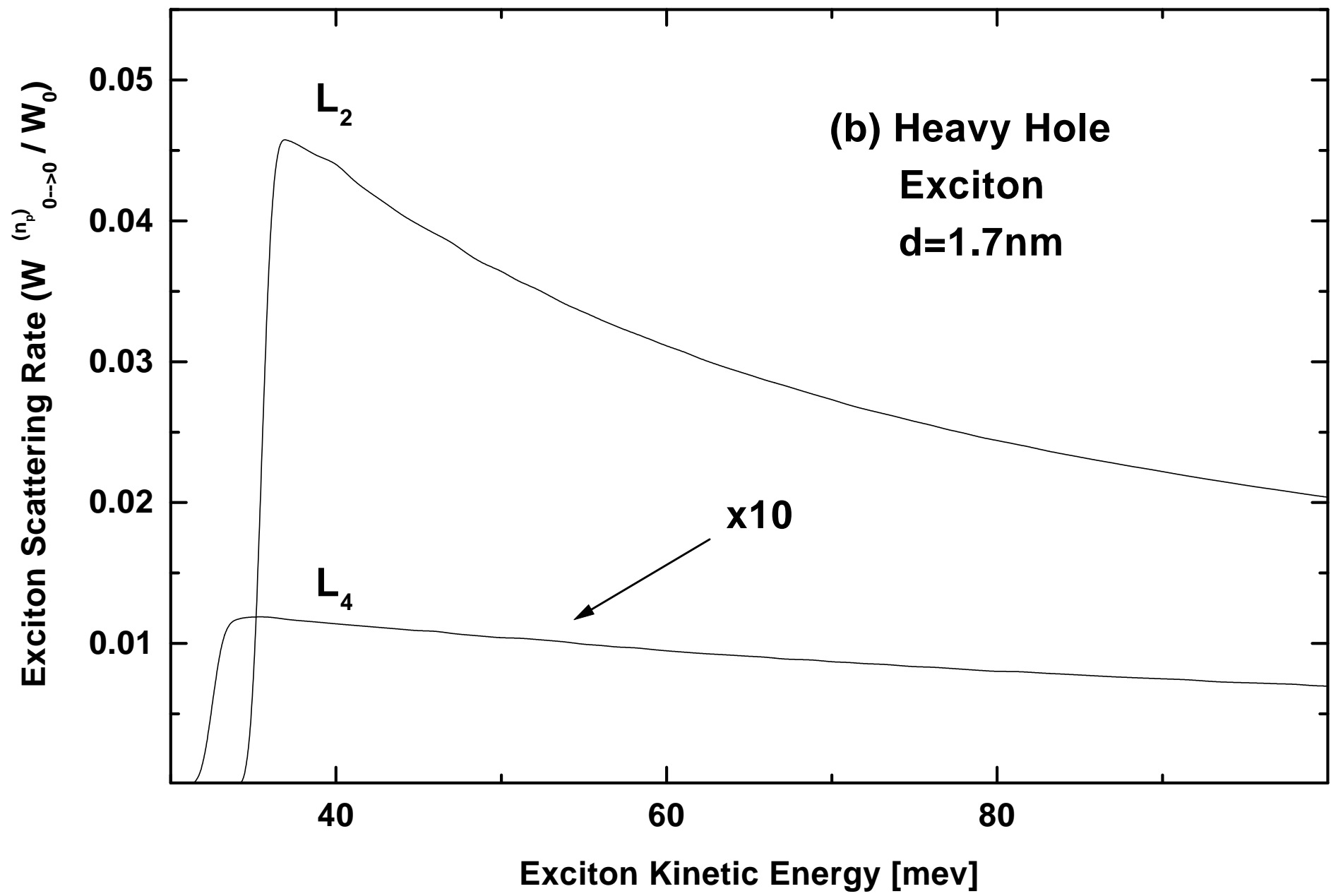
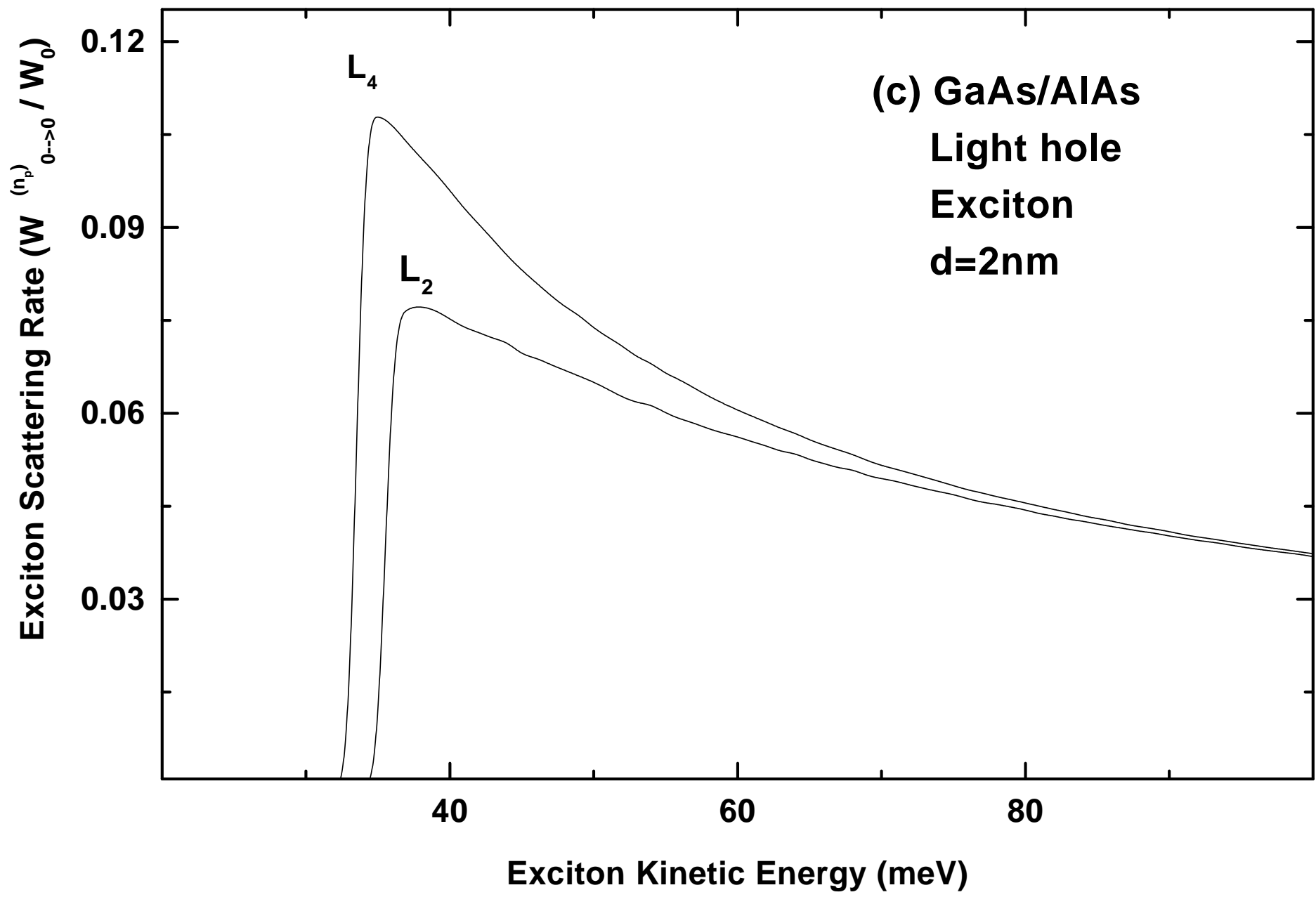
\end{document}